\documentclass[12pt]{iopart}

\usepackage{iopams}  
\usepackage{hyperref}
\hypersetup{colorlinks=true, linkcolor=blue, citecolor=red, urlcolor=blue}

\begin{document}

\title{\vspace{-3truecm}\\ A critical survey of twisted spectral triples \\beyond the Standard Model} 
 \author{Manuele Filaci}
 \address{University of Krakow, Institute fo Physics, Jagiellonian
   University\\ prof. Stanis\l{}awa \L{}ojasiewicza 11, 30-348 Krakow, Poland}
 \ead{manuele.filaci@uj.edu.pl}
\author{Pierre Martinetti}
 \address{Universit\`a di Genova (dpto di matematica) \& INFN, \\via
   Dodecaneso, 16146 Genova, Italia}
\ead{martinetti@dima.unige.it}



\newcommand{\ds}{{\slash\!\!\!\partial}}
\newcommand{\A}{{\mathcal A}}
\newcommand{\B}{{\mathcal B}}
\newcommand{\F}{{\mathcal F}}
\newcommand{\E}{{\mathcal E}}
\newcommand{\Asm}{{\mathcal A_\text{SM}}}
\newcommand{\D}{{\mathcal D}}
\newcommand{\HH}{{\mathcal H}}
\newcommand{\M}{{\mathcal M}}
\newcommand{\J}{\mathcal{J}}
\newcommand{\R}{\mathcal{R}}
\newcommand{\C}{{\mathbb C}}
\newcommand{\HHH}{{\mathbb H}}
\newcommand{\cinf}{{C^\infty\!(\M)}}
\newcommand{\I}{\mathbb I}
\newcommand{\dmu}{\text{d}\upmu}
\newcommand{\SU}[1]{SU\left(#1\right)}
\newcommand{\red}[1]{{\color{red}{#1}}}
\newcommand{\pert}{{\mathbb P}\mathrm{ert}}

\begin{abstract}
We review the applications of twisted spectral triples to the Standard
Model. The initial motivation was
to generate a scalar field, required to stabilise the elec\-troweak
vacuum and fit the Higgs mass, while respecting the first-order
condition. Ultimately,  it turns out that the truest interest of the twist lies in a
new -- and unexpected --  field of $1$-forms, which is related to the
transition from Euclidean to Lorentzian signature.
\end{abstract}



\section{Introduction}

From the pioneering work of \cite{Dubois-Violette:1989fk}
till the full formalism of Connes \cite{Connes:1994kx}, noncommutative geometry provides a
unified description of 
the Lagrangian of the Standard Model of
  fundamental interactions (electromagnetism, weak and strong
  interactions) \cite{Connes:1990ix}\cite{Chamseddine:1996ly}\cite{Chamseddine:1996kx};
minimally coupled to the Einstein-Hilbert action of
    General Relativity \cite{Connes:1996fu};
including right handed neutrinos \cite{Chamseddine:2007oz};
where the Higgs boson comes out naturally on the same
  footing as the other bosons, i.e. as the local expression of a
    connection $1$-form.



The approach works very well on Riemannian manifolds. The
generalisation to pseudo-Riemannian geometry, in particular Lorentzian manifolds,
is 
 far from obvious (there are various attempts in this direction,
 see for instance
 \cite{Barrett:2007vf}\cite{Besnard:2020ab}\cite{Franco:2012fk}\cite{Dungen:2015aa}\cite{Bochniak:2018aa}
 and reference within).

In addition, noncommutative geometry offers possibilities to go beyond
the Standard
Model, by modifying the rules of the game in various ways: one may
 enlarge the space of fermions
\cite{Stephan:2009fk,Stephan:2013ab}, 
or  get rid of the  \emph{first-order condition} (defined below)  \cite{Chamseddine:2013fk,Chamseddine:2013uq},
modify the real structure (also defined below)
  \cite{Brzezinski:2018aa,T.-Brzezinski:2016aa}, switch to
  non-associative geometry \cite{Boyle:2014aa, Boyle:2019ab}, use some
  structure of Clifford bundle in order to modify
  some of
  the mathematical requirements defining a noncommutative geometry
\cite{Dabrowski:2017aa}. For a recent review of ``beyond Standard
Model'' propositions in the framework of noncommutative geometry, see
\cite{Chamseddine:2019aa}. 

Here we focus on another class of variations around Connes' initial
model, obtained by twisting the noncommutative geometry
by an algebra automorphism \cite{Devastato:2013fk}\cite{buckley}\cite{Martinetti:2019aa}.

All the possibilities above 
 but the first are minimal extensions of the~Standard~Model, in that they
yield an extra scalar field $\sigma$ -- suggested
by particle physicists~to stabi\-lize the electroweak vacuum -- but do
not touch the fermionic content.~The novelty~of the twist is to generate another additional
piece: a new field of $1$-forms, which suprisingly turns out to
  be related to the transition from Euclidean to Lorentzian signature
  \cite{Devastato:2018aa}. In~particu\-lar, in the example of electrodynamics, this field identifies
with the (dual) of the $4$-momentum vector in Lorentzian signature,
even though one started with a Riemannian manifold \cite{Martinetti:2019aa}.

All this is explained as follows. In the next section we begin by
some recalling on the spectral description of the
Standard Model \cite{Chamseddine:2007oz}. We stress the process of
fluctuation of the metric, which is the way to generate bosonic fields
in noncommutative geometry by turning the constant parameters of the
model into fields.

In
section \ref{sec:grand-algebra-beyond} we describe the model of grand algebra
developed in \cite{Devastato:2013fk}, which aimed at generating the
extra scalar field $\sigma$, while  respecting the first-order condition. The
idea is to start with an algebra
bigger than the one of the Standard Model, in order to have more
``space'' to generate bosonic fields by fluctuations of the metric.
This model does indeed generate the
expected field $\sigma$, by letting the Majorana mass of the neutrinos
fluctuate. Even though the first-order condition associated with this
Majorana mass is preserved, the problem moves to the free Dirac
operator: not only does the latter break the first-order condition,
but its commutator with the algebra is unbounded, in contradiction
with 
the very definition of spectral triple. This kind of problem is
typically solved by twisting the spectral triple in the sense of Connes
and Moscovici~\cite{Connes:1938fk}. A twisting of the grand algebra
down to the Standard Model has been worked out in \cite{buckley}, but
we show in \S \ref{sec:twist-grand-algebra} that this does not define
stricto sensu a twisted
spectral triple, for only the part of the algebra relevant for
the extra scalar field has been twisted. 

Applying the twist to the whole algebra suggests a general procedure to twist any
graded spectral triple, as recalled in  section
\ref{sec:minimal-twist}. It consists in  doubling the algebra one is
beginning with, rather
than finding it from the reduction of  a bigger algebra. Such a
``twisting up'' procedure has been studied in a couple of papers
\cite{Lett.}\cite{Landi:2017aa}. There is some freedom in the
construction, especially in the choice of the twisting operator whose eigenspaces
determine the representation of the doubled algebra.  By choosing the grading as the
twisting operator, one automatically satisfies the twisted first-order
condition. However, when applied to the spectral triple of the Standard
Model, this twist-by-grading does not generate any extra scalar field.
Some
preliminary results, mentioned in \S \ref{sec:twist-fluct-with}, indicate that this is a
general feature of the
twisting-up procedure:  the twisted first-order condition and the
extra scalar field are mutually exclusive. Hence the necessity to
adapt to the twisted case the fluctuations without first-order
condition introduced in \cite{Chamseddine:2013fk}. This has been done in
\cite{Martinetti:2021aa} and is summarised in \S \ref{sec:twist-fluct-with}.

Section  \ref{sec:twist-loretnz} deals with what might  be the
truest interest of the twist, namely the~unex\-pected field of
$1$-forms arising from the twisted fluctuation. In the example of electrodynamics
\cite{Martinetti:2019aa},\cite{Dungen:2011fk},  this field identifies with the dual of the
$4$-momentum in Lorentzian signature,
 even though one started with
a Riemannian spectral triple.

\newpage

\section{The spectral description of the Standard Model}
\label{sec:spectr-descr-stand}

We begin with the definition of spectral triple, which is the central tool
in Connes' noncommutative geometry,  emphasising how the bosonic
fields -- including the Higgs field -- are obtained as
connection $1$-forms, through the process of 
\emph{fluctuation of the metric}. We then summarise the spectral description of
the Standard Model.

\subsection{Spectral triple}

A spectral triple \cite{Connes:1994kx} consists of an algebra
$\A$ acting on a Hilbert space $\HH$ together~with a selfadjoint
operator $D$ with compact resolvent, whose  
commutator 
$[D,a]$
is bounded~for any $a\in\A$.
\noindent It is graded if it comes with a selfadjoint operator
$\Gamma$ on $\HH$~which
squares to the identity operator $\I$, anticommutes with $D$ and
commutes with the algebra.
A spectral triple is real \cite{Connes:1995kx} if in addition there
is an antilinear operator $J$ on $\HH$~satisfying 
\begin{equation}
  J^2=\epsilon\I,\quad JD=\epsilon' DJ,\quad J\Gamma=\epsilon''
  \Gamma J
\end{equation}
 where $\epsilon, \epsilon', \epsilon''=\pm 1$
 define the
 $KO$-dimension $k\in\left[0,7\right]$. This
 real structure implements a map $a\to a^\circ :=  Ja^*J^{-1}$
 from $\A$ to the opposite
 algebra $\A^\circ$.
 This yields a right action of $\A$ on $\HH$,
$\psi a := a^\circ \psi$, 
 which
 is asked to commute with the left action
 \begin{equation}
   \label{eq:7}
   [a, Jb^*J^{-1}] = 0 \quad\quad\forall a \in \A \quad\textrm{(order zero condition).}
 \end{equation}
There is also a first-order condition \cite{Connes:1996fu},
  \begin{equation}
[[D,a], Jb^*J^{-1}] = 0\quad\quad\forall a,b \in \A
\label{eq:8}
\end{equation}
which is there to guarantee that the operator $D$ be a first-order
  differential operator.

All these properties are satisfied by the triple
\begin{equation}
(\cinf,\;  L^2(\M, S),\;  \ds)
\label{eq:13}
\end{equation}
where $\cinf$ is the (commutative)  algebra of smooth functions
on a closed Riemannian manifold $\M$ of dimension $m$, acting by multiplication on the
Hilbert space $L^2(\M, S)$ of square-integrable spinors on $\M$, and
\begin{eqnarray}
  \label{eq:2407}
  \ds=-i\sum_{\mu=1}^m \gamma^\mu(\partial_\mu + \omega_\mu),\quad\textrm{ with } \quad
  \gamma^\mu\gamma^\nu+\gamma^\nu\gamma^\mu = 2g^{\mu\nu}\I
\end{eqnarray} 
is the Dirac operator ($\omega^\mu$ is the spin connection,
$\gamma^\mu$ the Dirac matrices and $g_{\mu\nu}$ the Riemannian metric
on $\M$, all given in local coordinates).  For $m$ even, this spectral triple has grading the
product of the Dirac matrices (in dimension $4$, the matrix
$\gamma^5$) and real structure $\cal J$ the charge conjugation operator. Adding other conditions
\cite{connesreconstruct} (which are satisfied by the triple~(\ref{eq:13})),
one gets \emph{Connes' reconstruction theorem}, that  extends Gelfand duality
(between compact topological spaces and $C^*$-commutative algebras) beyond
topology. Namely, given any real
spectral triple $(\A, \HH,  D)$ satisfying these
conditions, with $\A$ commutative, then there exists a closed
Riemannian manifold $\M$ such that  
$\A\simeq \cinf$.

A \emph{noncommutative geometry} is then defined as
a spectral triple $(\A, \HH, D)$ where $\A$ is  non (necessarily) commutative.
This gives access to new geometrical objects, that can
be intended as  ``spaces'' whose algebra of functions $\A$ is
not commutative.
\newpage
\subsection{Connection}

Take a  gauge theory with gauge group $G$. From a mathematical point
of view,  the fermionic fields form sections of  a $G$-bundle $\E$
over the spacetime $\M$,
while the bosonic fields are described as connections on $\E$.

In noncommutative geometry the spacetime $\M$ is substituted by a
spectral triple  $(\A, \HH, D)$, where
$\A$ plays the role of ``algebra of functions'' on the
noncommutative space. To understand what plays the role of a gauge
bundle, recall that the set of sections of any bundle on a
manifold $\M$ forms a finite
projective $C^\infty(\M)$-module. Conversely, by Serre-Swan theorem,  any such module actually is the
module of sections of a bundle on $\M$. That is why, in
noncommutative geometry,  the role of gauge bundles is played  by 
finite projective $\A$-modules $\E$. 

Connections on these modules are, by definition,  objects associated with a
derivation. Recall that a derivation $\delta$ on an algebra $\A$ is a map from $\A$ to some $\A$-bimodule
$\Omega$  satisfying the Leibniz rule
\begin{equation}
\delta(ab)= a\delta(b) + \delta(a)b\qquad \forall a,b\in \A.
\label{eq:10}
\end{equation}
A connection on a (right) $\A$-module $\E$ associated with $\delta$
is a map
  from $\E$ to $\E\otimes_\A \Omega$ such that the following Leibniz
  rule holds,
\begin{equation}
\nabla (\eta a) = \nabla (\eta)a + \eta\otimes \delta(a) \quad
\forall \eta\in\E, \, a\in\A,\label{eq:9}
\end{equation}
where
\begin{equation}
\Omega= \left\{ \sum_i a_i\delta(b_i),\; a^i, b^i \in\A\right\}
\label{eq:11}
\end{equation}
is the
  $\A$-bimodule generated by the derivation $\delta$, while 
  $\nabla (\eta)a$  
is a shorthand notation for $\eta_{(0)}a\otimes \eta_{(1)}$, using Sweedler notations $\nabla\eta = \eta_{(0)}\otimes
\eta_{(1)}$ with $\eta_{(0)}\in\E$ and $\eta_{(1)}\in\Omega$.
\bigskip

\noindent{\bf Example}:
     The exterior derivative
    $d$ is a derivation on the algebra $\cinf$. It generates the
    $\cinf$-bimodule of section $s$ of the cotangent bundle,
 \begin{equation}
\Omega^1(\M):=\left\{\sum_i f_i dg_i \,\textrm{ with }\,  f_i,
    g_i\in\cinf\right\}.\label{eq:86}
  \end{equation}
A connection on the tangent bundle $TM$ associated with $d$
is a map  
    \begin{eqnarray}
      \nabla:& \Gamma^\infty(TM) &\to \Gamma^\infty(TM)\otimes\Omega^1(\M),\\
      &\partial_\nu &\mapsto \Gamma_{\mu\nu}^\rho \partial_\rho\otimes dx^\mu,
    \end{eqnarray}
where $\Gamma^\infty(TM)$ denotes the set of smooth sections of $TM$. One  retrieves the usual notion of connection, as a map from $\Gamma^\infty(TM)\times
\Gamma^\infty(TM)$ to $\Gamma^\infty(TM)$ 
as
    \begin{equation*}
\nabla: (\partial_\eta,\partial_\nu) \mapsto
 \nabla_\eta\partial_\nu :=   
    \Gamma_{\mu\nu}^\rho \partial_\rho\otimes_\cinf\langle
    dx^\mu,\partial_\eta\rangle \simeq \langle
    dx^\mu,\partial_\eta\rangle\Gamma_{\mu\nu}^\rho \partial_\rho=\Gamma_{\eta\nu}^\rho\partial_\rho,
\label{eq:12}
      \end{equation*}
where $\langle .\,, . \rangle$ is the $\cinf$-valued dual product between
the cotangent and the tangent bundles and the last equation is the
isomorphism between ${\cal E} \otimes_\cinf\cinf$ and $\cal E$.

\subsection{Fluctuation of the metric}

To understand when two
algebras are ``similar'', a relevant notion is  \emph{Morita equivalence}. This
is more flexible than isomorphism of algebras for, roughly speaking, two algebras $\A$ and $\B$ are
Morita equivalent if they have the same representation
theory. Technically, it means that there exists an Hermitian finite
projective $\A$-module $\E$ such that 
${\cal B}$ is isomorphic to the algebra $\textrm{End}_\A(\E)$ of
$\A$-linear, adjointable,  endormorphisms  of $\E$ (for details see
e.g. \cite{Rieffel:1982aa} or \cite{Landi:1997zm}).

The idea of fluctuation of the metric comes from the following
question: how does one  export a real spectral triple $(\A, \HH, D)$ to a Morita
equivalent algebra $\cal B$ ? We recall the construction of
\cite{Connes:1996fu}, whose details may be found in
\cite{Connes:2008kx} and even more details in~\cite{Landi:2017aa}.

Assume 
$\E= \E_R$ is a right $\A$-module.
The algebra $\cal B$ acts on
$\HH_R := \E_R\otimes_\A \HH$
as 
\begin{equation}
b(\eta\otimes \psi ) =b\eta\otimes \psi \quad\quad \forall
b\in{\cal B},
\eta\in\E_R, \psi\in\HH.\label{eq:87}
\end{equation}
However, the  ``natural'' action of $D$ on $\HH_R$,
\begin{equation}
  D_R(\eta\otimes\psi):=\eta\otimes D\psi,
\end{equation}
 is not compatible with
the tensor product, for 
\begin{equation}
 D_R(\eta a\otimes \psi) - D_R(\eta\otimes a\psi) = -\eta\otimes
[D,a]\psi
\label{eq:15}
\end{equation}
has no reason to vanish. This is cured by equipping $\E_R$
with a connection $\nabla$  with value in the $\A$-bimodule of
generalised $1$-forms
\begin{equation}
\Omega^1_D(\A):=\left\{\sum_i  a_i [D, b_i],\; a_i,
  b_i\in\A\right\}
\label{eq:23}
\end{equation}
generated
by the derivation $\delta(.)=[D,.]$. Indeed, the following operator,
\begin{equation}
D_R(\eta\otimes\psi):= \eta\otimes D\psi +
(\nabla\eta)\psi\label{eq:16}
\end{equation}
is well defined on $\HH_R$, and selfadjoint as soon as $\nabla$
is an hermitian connection. Moreover one checks that the
commutator $[D_R, b]$ is bounded for any $b\in B$, so that $(\B, \HH_R,
D_R)$ is a spectral triple.
In particular, if one considers self-Morita equivalence, that is
${\cal B}=\E_R=\A$, then the operator (\ref{eq:16}) with $\nabla$
hermitian reads 
\begin{equation}
D_R=D + A_R\label{eq:17}
\end{equation}
with $A_R=A_R^*\in\Omega^1_D(\A)$ a selfadjoint generalised $1$-form.

A similar construction holds if one implements Morita equivalence with
a left module $\E_L$. Then $\HH_L = \HH\otimes_\A \E_L$ is a Hilbert
space and the operator
\begin{equation}
 D_L(\psi\otimes\eta):=D\psi\otimes\eta
   + (\nabla^\circ\eta)\psi\label{eq:18}   
\end{equation}
with $\nabla^\circ$ a
 connection  with value in the bimodule
$$\Omega^1_D(\A^\circ)= \left\{\sum_ia_i^\circ [D, b_i^\circ],\quad a_i^\circ,
b_i^\circ\in\A^\circ\right\}$$
is well defined on  $\HH_L$. For $\nabla^\circ$ hermitian, one obtains
a
spectral triple $({\cal B}, \HH_L, D_L)$. For self-Morita equivalence, one gets
\begin{equation}
D_L=D+ A^\circ  = D + \epsilon' J\, A_L\, J^{-1}\label{eq:19}
\end{equation}
with $A^\circ\in\Omega^1_D(\A^\circ)$ and
$A_L\in\Omega^1_D(\A)$. 

To take into account the real structure, one combines the two constructions,
using an $\A$-bimodule $\E$ to  implement Morita equivalence. For
self-Morita equivalence,  one then obtains
 the operator 
$D'= D + A_R + \epsilon' J\, A_L
J^{-1}$ acting on $\HH$. Requiring this operator to be selfadjoint and $J$ to be a
real structure amounts
to the existence of a generalised selfadjoint $1$-form
     $A\in\Omega^1_D(\A)$ such that
     \begin{equation}
 D' = D_A:= D + A + \epsilon' J \, A J^{-1}.\label{eq:22}
 \end{equation}

Then $(\A, \HH, D_A)$ is a real spectral triple.
The operator $D_A$ is called a covariant Dirac operator, and the substitution of $D$ with a $D_A$ is  a \emph{fluctuation of the metric}.
The name is motivated by the existing relation between the Dirac
operator and the metric. This relation is  obvious on a spin manifold, from the very
definition of the Dirac matrices ( $\gamma^\nu\gamma^\nu + \gamma^\nu\gamma^\mu=2g^{\mu\nu}$), and it still makes sense for an arbitrary
noncommutative geometry, via the definition of the spectral
distance \cite{Connes:1992bc}. On
a manifold, this distance gives back the geodesic distance
associated with the Riemannian structure of $\M$, while on an arbitrary
spectral triple it may be seen as a generalisation of the
Wasserstein distance of order $1$ in the theory of optimal transport
(cf \cite{dAndrea:2009xr,Martinetti:2016aa} and references therein). By exporting a noncommutative
geometry to a Morita equivalent algebra, one passes from $D$ to the covariant operator
$D_A$ and  modifies accordingly the spectral distance. In particular,
for the Standard Model, such a fluctuation provides a purely
metric interpretation to the Higgs field (which is one of the components of
the generalised $1$-form $A$, see below) \cite{Connes:1996fu,Martinetti:2002ij}.  The metric interpretation of the
other components of $A$ has been worked out in
\cite{Martinetti:2002ij,Martinetti:2008hl}, in relation with the
Carnot-Carath\'eodory distance in sub-Riemannian geometry.
 
\subsection{Gauge transformation}

A gauge transformation is a change of connection on
the Morita-equivalence bimodule~$\E$. In case of self-Morita
equivalence, it is implemented by the conjugate action
on $\HH$ of the group $U(\A)$ of unitaries element of $\A$
(i.e. $u\in\A$ such that $u^*u = uu^*=\I$):
\begin{equation}
\textrm{Ad}(u): \psi \to u\psi u^* = u (u^*)^\circ \psi = u J
uJ^{-1}\psi\quad\forall \psi\in\HH.\label{eq:14}
\end{equation}

This action maps the covariant Dirac operator $D_A$ to 
\begin{equation}
\textrm{Ad}(u)\, D_A \, \textrm{Ad}(u)^{-1} 
\label{eq:20}
\end{equation}
and one checks that this operator coincides with 
the operator $D_{A^u}$, defined as in (\ref{eq:22}) with 
\begin{equation}
A^u:= u[D, u^*] + uAu^*.\label{eq:21}
\end{equation}

This is the formula of transformation of the gauge potential in
noncommutative geometry, which generalises the usual one of gauge theories.

\subsection{Standard Model}
\label{sec:standard-model}

The spectral triple of the Standard Model \cite{Chamseddine:2007oz} is the product 
\begin{eqnarray}
\label{eq:2}
\hspace{-1.5truecm} 
\A = \cinf \otimes A_{F},\quad \HH = L^2(\M, S)\otimes H_{F}, \quad D= \ds\otimes
  \I_{96} + \gamma^5\otimes D_{F}
\end{eqnarray}
of the
spectral triple (\ref{eq:13})
of a $4$-dimensional  Riemannian closed spin manifold~$\M$ with a finite
 dimensional spectral triple

\begin{eqnarray}
\hspace{-1.5truecm}  A_{F}= \C \oplus {\mathbb H} \oplus M_3(\C), \quad
  H_{F}= \C^{96},\quad
D_{F}=\underbrace{\left(\begin{array}{cc}
D_0 &0_{48}\\
0_{48}&  D_0^\dagger \end{array}\right)}_{D_Y} 
+
\underbrace{\left(\begin{array}{cc}
0_{48}&D_R\\
D_R^\dagger &0_{48} \end{array}\right)}_{D_M}
\end{eqnarray}
where $\HHH$ is the algebra of quaternions and $M_3(\C)$ the algebra of
complex $3\times 3$ matrices.

The dimension of $H_F$ is the number of fermions in the Standard
Model (including right-handed neutrinos). Its entries are labelled by
a multi-index $\,C\,I\,\alpha\, n$
where 

  \begin{itemize}
  \item[$\bullet$] $C=0,1$ labels particles $(C=0)$ or anti-particles
    $(C=1)$;
  \item[$\bullet$] $I=0,\, i$ with $i=1,2,3$ is the lepto-colour
    index: it takes value $I=0$
    for a lepton and $I=1, 2, 3$ for a quark with its three possible
    colours;
  \item[$\bullet$] $\alpha=\dot{1},\dot{2},1, 2$ is the flavour index
    (with dot indicating the chirality):
    \begin{eqnarray}
      &\dot{1}=\nu_R, \; \dot{2}=e_R,\; 1=\nu_L,\; 2=e_L&\textrm{ for leptons ($I=0$)},\\
      &\dot{1}=u_R, \; \dot{2}=d_R,\; 1=q_L,\; 2=d_L&\textrm{ for
                                                      quarks } (I=i); \end{eqnarray}
\item[$\bullet$] $ n=1, 2, 3$ is the generation index.
  \end{itemize}

The details of the representation of $A_F$ is in
\cite{Chamseddine:2007oz}. The important point for our matter is that the
quaternions act only on the particle subspace of $H_F$ $(C=0)$, trivially on the lepto-colour
index $I$, and through their fundamental
representation on the last two flavour indices $\alpha$; whereas $M_3(\C)$ acts only on
antiparticle subspace of $H_F$ ($C=1$), trivially on the flavour
index $\alpha$
and through their fundamental representation
on the lepto-colour index $i$. The algebra $\C$ acts both on particles
together with the quaternions (but on the first two flavour indices),
and on antiparticles together with $M_3(\C)$ (on $I=0$). 

The grading of the finite dimensional spectral triple is the $96\times
96$ matrix $\Gamma_F$ with entries $+1$ on left particles/right antiparticles,
$-1$ on right particles/left antiparticles. The real structure is the
matrix $J_F$ that exchanges particles with antiparticles. The
spectral triple (\ref{eq:2}) is real, with grading  
$\Gamma=\gamma^5\otimes\Gamma_F$
and real structure $J = \mathcal{J}\otimes J_{F}$. 

In the particles/antiparticles indices, the Dirac operator $D_F$ of the finite dimensional spectral triple is the
sum of a block diagonal matrix $D_Y$ which contains the Yukawa
couplings of the fermions, the Cabibbo-Kobayashi-Maskawa mixing matrix for the quarks and
the Pontecorvo-Maki-Nakagawa-Sakata mixing matrix for the left-handed neutrinos, and a block off-diagonal  matrix
$D_M$ which contains the Majorana masses $k_R^n$, $n=1, 2, 3$ of the
right-handed neutrinos and the corresponding mixing matrix (notations are those of
\cite{M.-Filaci:2020aa}, they differ from the ones of
\cite{Devastato:2013fk} and \cite{buckley}).

The generalised $1$-forms (\ref{eq:23}) for a product of spectral
triples  (\ref{eq:2}) decompose as~\cite{D.-Kastler:1993aa}
\begin{equation}
  A = \gamma^5\otimes H -i\sum_\mu \gamma^\mu \otimes A_\mu
\end{equation}
 where
$H$ is a  scalar field on $\M$ with values in
 $A_F$, while  $A_\mu$ is  a $1$-form field on $\M$ with values in the Lie algebra of
the group  $U(A_F)$ of unitary elements of $A_F$ (differently said: a connection $1$-form on a $U(A_F)$-bundle on $T\M$). In particular, for the
 spectral triple of the Standard Model, one has
 \begin{equation}
U(A_F)= U(\C \oplus {\mathbb H} \oplus
 M_3(\C))\simeq U(1)\times SU(2)\times U(3),\label{eq:1}
 \end{equation}
which is reduced to
 the gauge group $U(1)\times SU(2)\times SU(3)$ of the Standard Model
 by imposing a unimodularity
 condition (which also guarantees that the model is anomaly free, see
 e.g \cite[\S 2.5]{Chamseddine:2007oz}). 

The action of this group on
 $\HH$ is a matrix whose components are the hypercharges of the fermions of the Standard
 Model \cite[Prop. 2.16]{Chamseddine:2007oz}.  This allows to identify the basis elements of $H_F$ with the
 $96$ fermions of the Standard Model, and the corresponding elements in
 $\HH$ with the fermionic fields. Moreover, the action of $A+ JAJ^{-1}$
 corresponds to the bosonic hypercharges, and allows to identify the components of $A_\mu$
 with the bosonic fields of the Standard Model
 \cite[Prop. 3.9]{Chamseddine:2007oz}. One also checks that (\ref{eq:21}) yields the
 expected gauge transformation.

The interpetation of the scalar field $H$ follows from the computation
of the \emph{spectral action}
\cite{Chamseddine:1996kx,Chamseddine:1996ly}, namely the asymptotic expansion $\Lambda\to\infty$ of
  $\textrm{Tr}\; f(\frac{D_A^2}{\Lambda^2})$
where $f$ is a smooth approximation of the characteristic function of the
interval $[0,1]$. One obtains 
the bosonic Lagrangian of the Standard Model
coupled with 
Einstein-Hilbert action in Euclidean signature, where $H$ is the Higgs field.
The coupling constants of the electroweak and strong interactions
satisfy the relation expected in grand unified theories, and are related
to the value at $0$ of the function $f$.

 The spectral action provides some relations between the parameters of the
 Standard Model at a putative unification
  scale. The  physical predictions are obtained by running down
 the parameters of the theory under the renormalisation group
 equation, taking these relations as initial conditions.
 Assuming there is no new physics between the unification scale and
 the electroweak scale, one finds a value for the Higgs mass around
$170 \textrm{ GeV}$, in disagrement with the measured value $125,1\textrm{ GeV}$.

However, for a Higgs boson with mass $m_H\leq 130 \,\textrm{Gev}$, the quartic
coupling $\lambda$ of the Higgs field becomes negative at high energy, meaning the
electroweak vacuum is meta-stable rather than stable
\cite{Degrassi:2012fk}.
This instability can be cured by a
new scalar field $\sigma$ which couples to the Higgs field.
In the spectral description of the Standard Model, such a field is obtained by turning into a field  the
neutrino Majorana mass~$k_R$ which appears in the off-diagonal
part $D_R$ of the finite dimensional Dirac operator $D_F$:
$$k_R\to k_R\sigma,$$
Furthermore, by
altering the~running of the parameters under the equations of the group
of renormalization,  this extra scalar field makes the computation of
the mass of the Higgs boson compatible
with its experimental value \cite{Chamseddine:2012fk}.

\section{Grand algebra beyond the Standard Model}
\label{sec:grand-algebra-beyond}

The point in the above is to justify the turning of the
constant $k_R$ into a field $k_R\sigma$. This cannot be obtained  by fluctuation
of the metric, for one checks that 
\begin{equation}
  \label{eq:5}
    [\gamma^5\otimes D_M, a]= 0 \quad\quad\forall a, b\in \A=\cinf\otimes A_{F}.
\end{equation}
In other terms, the constant $k_R$ is transparent under fluctuation. The
solution proposed in \cite{Chamseddine:2013fk} is to remove the
first-order condition. This gives more flexibility, and permits to
obtain the extra scalar field as a \emph{fluctuation without the first-order condition}. The latter is retrieved dynamically, by minimising the spectral
action \cite{Chamseddine:2013uq}. In this way the field $\sigma$ is the ``Higgs'' boson
associated with the breaking of the first-order condition.

\subsection{Grand algebra}
\label{sec:grand-algebra}

At the same time, an alternative process was described in
\cite{Devastato:2013fk} where one mixes the internal degrees of
freedom per generation of the finite dimensional Hilbert space $H_F$,
that is 
$\HH_F\simeq \C^{32}$, with the $4$ spinorial degrees of freedom of
$L^2(\M, S)$. This provides exactly the
$4\times 32= 128$ degrees of freedom required to represent the
``second next algebra'' in the classification of finite
dimensional spectral triples made in \cite{connesneutrino,
  Chamseddine:2008uq}. 

In this classification, the smallest algebra -- $\HHH\oplus  M_2(\C)$ --
is too small to
accomodate the Standard Model; the second smallest one --
$\A_{SM}= M_2(\HHH)\oplus  M_4(\C)$  -- reduces to the left-right
algebra $\A_{LR}= \HHH_L\oplus\HHH_R \oplus M_4(\C)$ by imposing 
the grading condition, which breaks to the algebra $\A_F$ of the Standard Model by the first-order
condition. The next one is $M_3(\HHH)\oplus  M_6(\C)$ and has not
found  any physical interpretation so far. Then comes the \emph{grand
  algebra}  \cite{Devastato:2013fk}
\begin{equation}
\A_G= M_4(\HHH)\oplus  M_8(\C).
\label{eq:26}
\end{equation}
It is too big to be represented on the Hilbert space $\HH_F$ in a way
compatible with the axioms of noncommutative geometry:  the latter require a
space of dimension $d=2(2a)^2$, where $a$ is the dimension of the quaternionic
matrix algebra. For $\A_{SM}$ one has  $a=2$,  which corresponds
to $d=2(2\cdot 2)^2=32$, that is the dimension of $\HH_F$. For the grand algebra $\A_G$, $a=4$ and one needs a space
four times bigger. 

This bigger space is obtained by allowing
$\cinf$ to act independently on the left and right components of
spinors. This permits to represent on $L^2(\M, S)\otimes \HH_F$  the algebra $\cinf\otimes\A_G$
-- viewed as functions on
$\M$ with value in $\A_G$ -- in such a way that
for any $a\in \cinf\otimes\A_G$ and $x\in\M$, then $a(x)\in\A_G$ acts
on $\HH_F$ in agreement with the classification of
\cite{Chamseddine:2008uq}. 

Within the tensorial notation of \S
\ref{sec:standard-model}, the components $M_4(\HHH)$ and $M_8(\C)$ of
the grand algebra are $2\times 2$ matrices $Q, M$ 
 with entries in $M_2(\HHH)$ and 
$M_4(\C)$ that  act on $\HH_F$ as
$\A_{SM}$. The difference with the spectral triple of the
Standard Model is that, once tensorised by $\cinf$,  the $2\times 2$
matrices $Q, M$ have
a non-trivial action on the spinorial degrees of freedom of $L^2(\M ,S)$.
We denote the latter by two indices: $s=l, r$ 
for the left/right components of spinors; $\dot s=\dot 0, \dot 1$ for
the particle/antiparticle subspaces. 

In \cite{Devastato:2013fk} one makes $\cinf\otimes M_4(\HHH)\ni Q$,
resp. $\cinf\otimes M_8(\C)\ni M$,  act non
trivially on the $\dot s$, resp $s$, index.
Omitting all the indices on
which the action is trivial,
 \begin{equation}
  \label{eq:32}
  Q=  \left(
\begin{array}{cc}
Q^{\dot 0 \beta}_{\dot 0 \alpha} & Q^{\dot 1 \beta}_{\dot 0 \alpha}\\
Q^{\dot 0 \beta}_{\dot 1 \alpha}&Q^{\dot 1 \beta}_{\dot 1 \alpha}
  \end{array}
\right) _{\dot s\dot t},\qquad   M=  \left(
\begin{array}{cc}
M^{r J}_{r I} & M^{l J}_{rI}\\
M^{r J}_{l I}&M^{l J}_{l I}
  \end{array}
\right) _{ s t},
\end{equation}
where $\beta$, $J$, $t$ and $\dot t$  are summation 
indices within the same range as $\alpha$, $I$, $s$, $t$ (the
indices after the closing parenthesis are those labelling the matrix entries).

Since $\gamma^5$ acts non trivially on
the spinorial chiral index, the grading condition forces $M$ to be diagonal in the
$st$ indices: $M^{l
  J}_{rI}=M^{l J}_{lr}=0$. Since $\Gamma_F$ is non trivial only in the flavour index $\alpha$,
in which the remaining entries $M^{lJ}_{lI},
M^{r J}_{rI}\in M_4(\C)$ act
trivially, the grading 
does not induce any further breaking in the complex sector. On the
contrary, since $\gamma^5$ is trivial in the $\dot s$ index but quaternions
act non trivially on the $\alpha$ index, the grading forces $Q$
to be diagonal in the flavour index, with components ${Q_L}_{\dot
  s}^{\dot t}$, $ {Q_R}_{\dot s}^{\dot t}\in \cinf\otimes M_2(\HHH)$ acting on
the left/right subspaces of the internal Hilbert space~$\HH_F$. In other terms, the grading condition breaks the grand algebra in
\begin{equation}
  \label{eq:29}
  \A'_G=\left(M_2(\HHH)_L\oplus M_2(\HHH)_R\right)\oplus
  \left(M_4(\C)_l\oplus M_4(\C)_r\right).
\end{equation}
To guarantee the  first-order condition for the Majorana
component 
$\gamma^5\otimes D_R$ of the Dirac operator, a solution is to further
break $\A'_G$ to 
\begin{equation}
\label{eq:41}
  \A''_G=\left(\HHH_L \oplus \HHH'_L \oplus \C_R \oplus \C'_R\right) \oplus
 \left( \C_l\oplus M_3(\C)_l\oplus  \C_r\oplus M_3(\C)_r\right)
\end{equation}
 with $\C_R= \C_r=\C_l$.
In the first term, the unprimed algebras act on the particle
subspace $\dot s = \dot 0$, while the primed ones act on the
antiparticle subspace $\dot s=\dot 1$. A fluctuation of the metric of $\gamma^5\otimes D_R$ then
yields an extra scalar  field $\sigma$, which lives in the difference
between $\C_R$ and $\C'_R$, and fixes the Higgs mass as expected \cite{Devastato:2014fk}.
In this grand algebra model, the fermionic content is not altered, since the total
 Hilbert space $\HH$ is untouched. One also checks the order zero condition.

The first-order condition for the free part $\ds\otimes\I$ of the
Dirac operator forces the components acting on the chiral spinorial
index to be equal, as well as those acting on~the \linebreak
particle/antiparticle index, meaning $\HHH'_L= \HHH_L$, $\C'_R=\C_R$
and $M_3(\C)_l=M_3(\C)_r$. Thus $\A''_G$ reduces to 
$\HHH_L \oplus \C_R \oplus M_3(\C)$,
namely the algebra of the Standard Model.
 The field~$\sigma$ \linebreak
thus appears  as the
Higgs field related to the breaking of the first-order condition by
$\ds\otimes\I$, whereas in
\cite{Chamseddine:2013fk} it is related with the first-order condition for $\gamma^5\otimes D_R$. By enlarging the algebra,
one has moved the symmetry
breaking from the internal space to~the~manifold.

However, the price to pay for a non trivial action on spinors
is the unboundedness of  the commutator of $\ds\otimes\I$ with the grand
 algebra: for $a=f\otimes m\in
 \cinf\otimes\A_G$~one~has 
 \begin{equation}
   \label{eq:27}
   [\ds\otimes \I, a] =   [\ds, f]\otimes m= -i[\gamma^\mu\partial_\mu,
   f]\otimes m -i[\gamma^\mu\omega_\mu,  f]\otimes m.
 \end{equation}
The second term is always bounded. By the Leibniz rule, the first
one is
\begin{equation}
  \label{eq:28}
  -i[\gamma^\mu, f]\partial_\mu - i\gamma^\mu(\partial_\mu f),
\end{equation}
which is bounded  iff the component $\partial_\mu$
vanishes. Since the only matrix that commutes with all the Dirac
matrices is the identity matrix, the commutator 
(\ref{eq:27}) is bounded if and only if $f$ acts on $L^2(\M, S)$ as a
multiple of the identity matrix, that is on the
same way on the left and right components of spinors. 
 
 \subsection{Twisted spectral triples}
\label{sec:twist-spectr-tripl}

Mixing the spinorial and internal degrees of freedom of
the Hilbert space $\HH$ -  in order to represent an algebra bigger than
the one of the Standard Model -  turns out to be incompatible with the very definition
of spectral triple. As explained at the end of the preceding section, this does not depend on the details of
the representation: as soon as the grand algebra acts
non trivially on spinors,  then the commutator with the free part of the
Dirac operator is unbounded \cite{Martinetti:2014aa}, no matter if the representation is
(\ref{eq:32}) or not.

The unboundedness of the commutator is the kind of problems adressed by  Connes and Moscovici when
they define \emph{twisted spectral triples} in
\cite{Connes:1938fk}. Their motivation had little to do with physics, but were purely mathematical (building
spectral triples with type III algebras).
 Given a triple $(\A, {\cal H}, D)$, instead of asking the commutators
  $[D,a]$ to be bounded, one asks the boundedness of the
 twisted commutators
 \begin{equation}
 [D, a]_\rho := Da - \rho(a) D 
 \label{eq:31}
 \end{equation}
for some fixed
   automorphism $\rho\in \textrm{Aut}(\A)$. 

The whole process of fluctuation of the metric has been adapted to the
twisted case  in \cite{Lett., Landi:2017aa}. One
obtains the  covariant Dirac operator
\begin{equation}
 D_{A_\rho}:= D+ A_\rho+J \, A_\rho \, J^{-1} \label{eq:24}
 \end{equation}
  where $A_\rho$ is an element of the set of  twisted  $1$-forms
 \begin{equation}
\Omega^1_D(\A, \rho):=\left\{\sum_i  a_i[D, J b_i^* J^{-1}]_{\rho^\circ}, \,
  a_i, b_i\in\A\right\}\label{eq:3}
\end{equation}
with $\rho^\circ:=\rho(a^*)^\circ$ is the automorphism of the opposite
algebra $\A^\circ$ induced by $\rho$. There is also twisted version of the first-order condition \cite{buckley,Lett.}
\begin{equation}
 [[D,a]_\rho,\, Jb^* J^{-1}]_{\rho^\circ}=0 \quad \forall a, b\in \A.
 \label{eq:4}
 \end{equation}
A gauge transformation is implemented by the twisted action of the
operator $\textrm{Ad}u$ (\ref{eq:20}),
\begin{equation}
\rho(\textrm{Ad}u)\,D_{A_\rho}\,\textrm{Ad}u^{-1},
\label{eq:58}
\end{equation}
with $\rho(\textrm{Ad}u): = \rho(u) J \rho(u)J^{-1} $ .
Such a transformation maps $ D_{A_\rho}$ to  $D_{A_\rho^u}$
where
\begin{equation}
  \label{eq:59}
  {A_\rho^u}= \rho(u) [D, u^*]_\rho + \rho(u) A^\rho u^*.
\end{equation}
This is  the twisted version of the gauge transformation (\ref{eq:21}).

\subsection{Twisting the grand algebra}
\label{sec:twist-grand-algebra}

To resolve the unboundedness of the commutator
arising in the grand algebra model, the idea is to find an
automorphism of $\cinf\otimes\A_G$ such that the twisted
commutator (\ref{eq:31}) of any element $(Q, M)\in\cinf\otimes\A_G$ with $\ds\otimes\I$ be bounded. This must be true
in particular for $(Q, 0)$ and $(0,M)$. Repeating the computation
(\ref{eq:27}) (\ref{eq:28}),
and taking into account only the spinorial indices $s, \dot s$ (since $\ds\otimes \I$ acts as the identity on all the other indices,
the corresponding sector of the algebra must be invariant under the automorphism, for
$\I a - \rho(a)\I=0$ iff $a=\rho(a)$), one finds that $\rho$ should be
such that
\begin{equation}
  \label{eq:25}
  \gamma^\mu Q - \rho(Q)\gamma^\mu =0\; \textrm{ and } \;   \gamma^\mu M -
  \rho(M)\gamma^\mu =0 \quad \forall\mu =1, ..., \textrm{dim} \,\M
\end{equation}
 for any $Q\in M_4(\HHH)\otimes\cinf$ and  $M\in M_8(\C)\otimes
 \cinf$. By easy computation, using the explicit form of the $\gamma$ matrices in the chiral basis,
\begin{equation}
  \label{eq:35}
  \gamma^\mu=
 \left( \begin{array}{cc}
    0_2 & \sigma^\mu \\
\bar\sigma^\mu & 0_2
  \end{array}\right)_{st} \quad \sigma^\mu=\left\{\I, \sigma^i\right\},\bar\sigma^\mu=\left\{\I, i\sigma^i\right\},
\end{equation}
where $\sigma^i$ are the Pauli matrices, one checks that any two $4\times 4$ complex matrices $A, B$ such that
$A\gamma^\mu = \gamma^\mu B$ for any $\mu$ are necessarily of the form
\begin{equation}
  \label{eq:36}
  A= \left( \begin{array}{cc}
    \lambda \I_2& 0_2 \\
0_2 & \lambda'\I_2
  \end{array}\right) \qquad   B= \left( \begin{array}{cc}
    \lambda' \I_2& 0_2 \\
0_2 & \lambda\I_2
  \end{array}\right) \quad\textrm{for some } \lambda, \lambda'\in\C. 
\end{equation}
Thus 
(\ref{eq:25}) implies that both $M$ and $Q$
must be  trivial in the $\dot s$ index, diagonal in the chiral
index $s$, with $\rho$  the autormorphism that exchanges the left
and right components.
 Therefore the representation (\ref{eq:32}) of the grand
algebra is not suitable to build a twisted spectral triple. 

In order to find a good representation, remember 
that the field $\sigma$ has its origin in the two copies $\C_R$,
$\C'_R$ of $\C$ in $\A''_G$
(\ref{eq:41}), which come from the non-trivial action of $\cinf\otimes
M_4(\HHH)$ on the
$\dot s$ index. Since the latter is no longer allowed, it seems
natural to make $\cinf\otimes M_4(\HHH)$ act non trivially on the chiral index
$s$. 
  On the contrary, 
the complex sector plays no obvious role in the generation of the
field $\sigma$, so one lets $\cinf\otimes M_8(\C)$ act trivially on
both the $s$, $\dot s$ indices.
 On the other indices, the action of   $M_4(\HHH)$,  $M_8(\C)$  is as in the Standard~Model.
The grading condition now breaks $M_4(\HHH)$ to $ \HHH_L^l \oplus \HHH_L^r
\oplus \HHH_R^l \oplus \HHH_R^r$ but leaves $M_8(\C)$ untouched. Reducing the
latter ``by hand'' to $M_4(\C)$, one gets the algebra \cite{buckley}
\begin{equation}
  \label{eq:6}
 {\cal B'}=  \HHH_L^l \oplus \HHH_L^r \oplus \HHH_R^l \oplus \HHH_R^r\oplus M_4(\C).\end{equation}
Let $\rho$ be the automorphism of $\cinf\otimes\cal B'$ that flips the
chiral spinorial
degrees of freedom,
\begin{equation}
  \label{eq:37}
  \rho(q_L^l, q_L^r, q_R^l,q_R^r, {\rm m}):=(q_L^r, q_L^l, q_R^r, q_R^l, {\rm m})
\end{equation}
where each of the $q$ is a quaternionic function with value  in its respective copy of $\HHH$ and
${\rm m}\in \cinf\otimes M_4(\C)$. Then 
\begin{equation}
  \label{eq:39bis}
(\cinf\otimes {\cal B'},  L^2(\M, S)\otimes \C^{32},\; \ds\otimes \I)
\end{equation}
is a twisted spectral triple which satisfies the  first-order condition \cite[Prop. 3.4]{buckley}.

Regarding the Majorana Dirac
operator, let us consider the subalgebra of ${\cal B}'$
\begin{equation}
  \label{eq:44}
  \tilde{\cal B}= \HHH_L^l \oplus \HHH_L^r \oplus \C_R^l \oplus \C_R^r\oplus
   (\C\oplus M_3(\C)).
\end{equation}
Given two of its elements $(q_L^l, q_L^r, c_R^l,
c_R^r, c, m)$, $(r_L^l, r_L^r, d_R^l, d_R^r, d, n)$ with $c, d, c_R^l,
c_R^r, d_R^l, d_R^r$ complex functions, $q_L^l, q_L^r, r_L^l,
r_L^r$ quaternionic functions and $m, n$ functions with values in $M_3(\C)$,
denoting $\pi'$ the representation of ${\cal B}'$ in the spectral
triple (\ref{eq:39bis}), one finds that 
\begin{equation}
  \label{eq:42}
[\gamma^5\otimes D_R,  \pi'(q_L^l, q_L^r, c_R^l, c_R^r, c, m)]_{\rho},
 \pi'(r_L^l, r_L^r, d_R^l, d_R^r, d, n)]_{\rho}
\end{equation}
vanishes as soon as $c=c_R^l$ and $d=d_R^l$ (or $c=c_R^r$ and
$d=d_R^r$). 
\newpage

In \cite{buckley}, this was improperly
interpreted  as a breaking of  ${\cal B'}$ to
\begin{equation}
  \label{eq:40}
  {\cal B}= \HHH_L^l \oplus \HHH_L^r\oplus \C_R^l \oplus
  \C_R^r \oplus M_3(\C).
\end{equation}
acting as $\tilde{\cal B}$ with $\C=\C_R^l$, namely the representation
$\pi$ of $\cal B$ is
\begin{equation}
  \label{eq:45}
  \pi(q_L^l, q_L^r, c_R^l, c_R^r, m):= \pi'(q_L^l, q_L^r, c_R^l, c_R^r, c_R^l,m).
\end{equation}
But $\rho$ exchanges the
left/right components  in the quaternionic sector only, so that 
\begin{equation}
  \label{eq:46}
  \pi'(\rho(q_L^l, q_L^r, c_R^l, c_R^r, c_R^l,m))= \pi'(q_L^r, q_L^l, c_R^r, c_R^l, c_R^l,m)
\end{equation}
is not the representation (\ref{eq:45}) of any element in $\cinf\otimes {\cal
  B}$ (the latter requires the identification of the first and third
complex functions, whereas in (\ref{eq:46}) the second and third are
identified), unless $c_R^r = c_R^l$. 
This means that the breaking from ${\cal B}'$ to ${\cal B}$ is not
compatible with the twist unless $\C=\C_R^l$ identifies with $\C_R^r$.
In that case, ${\cal B}'$ actually breaks to
$\HHH_L^l \oplus \HHH_L^r\oplus \C\oplus
 M_3(\C)$.
This algebra contains only one copy of $\C$ and so does not generate
the field $\sigma$ by twisted fluctuation of $\gamma^5\otimes D_R$. 

In
other terms,  the model developed in \cite{buckley} does not allow to generate the
 extra scalar field while preserving the first-order condition (even in a
 twisted form), as opposed to what was claimed. 
The error is due to not noticing that the reduction from
$\tilde{\cal B}$ to $\cal B$, imposed by the twisted first-order
condition of the Majorana Dirac operator, is not invariant under the twist.
So it does not make sense to try to build a spectral triple with
$\cinf\otimes\cal B$.

Nevertheless all the expressions computed in \cite{buckley}  of the
form
\begin{equation}
  \label{eq:51}
  T\pi'(a) - \pi'(\rho(a))T
\end{equation}
for $T=\ds\otimes\I$ or
$\gamma^5\otimes D_R$ are algebraically correct.  The point is that they are twisted
commutators  (\ref{eq:31}) for  $a$ in $\cinf\otimes\tilde{\cal B}$,
but not for $a$ in $\cinf\otimes{\cal B}$. Indeed, although (\ref{eq:46})  does define a representation of
$\cinf\otimes{\cal B}$,
\begin{equation}
  \label{eq:43}
  \hat\pi(q_L^l, q_L^r, c_R^l, c_R^r, m):= \pi'(q_L^r, q_L^l, c_R^r, c_R^l, c_R^l,m),
\end{equation}
there  is no automorphism
$\eta$ of $\cinf\otimes{\cal B}$ such that $\hat\pi$ would equal
$\pi\circ\eta$. 
What the results of  \cite{buckley}  show is that starting with the twisted spectral triple
\begin{equation}
(\cinf\otimes\tilde{\cal B}, L^2(\M, S)\otimes \HH_F, \ds\otimes \I + \gamma^5\otimes D_F),
\label{eq:52}
\end{equation}
whose Majorana part violates the twisted first-order condition,
then 
a twisted fluctuation of the Dirac operator by the subalgebra
$\cinf\otimes\B$ yields the field $\sigma$. Minimising the spectral action
(suitably generalised to the twisted case) breaks the algebra to the
one of the Standard Model, which satisfies the first-order
condition.

As noticed at the end of \cite{Lett.}, an alternative way to interprete (\ref{eq:51})
for $a$ in $\cinf\otimes\cal B$ is to view it
as a twisted commutator for the represented algebra. Namely defining
 the inner
automorphism 
 $\alpha_U (B):= UBU^*$ 
of ${\cal B}({\cal H})\supset \cal B$ that exchanges the $l, r$
components in the particle sector $C=0$ of $\HH_F$ (it is implemented
by the unitary $U=\gamma^0 \otimes P + \I \otimes (\I-P)$ with $P$ the projection on the particle
subspace of $\HH_F$), then   (\ref{eq:51}) reads as 
\begin{equation}
 T\pi(a) - \alpha_U(\pi(a))T \quad \textrm{ for } \quad a\in \cinf\otimes\cal B.\label{eq:33}
 \end{equation}
It is not yet clear whether this observation is
of interest.

\subsection{Twisting down}
\label{sec:twisting-down}

In the light of the preceding section, the conclusion of
\cite{buckley} should be corrected: twisted
spectral triples do resolve the unboundedness of
the commutator arising in the grand algebra model, but the extra scalar field breaks the
first-order condition, even in its twisted form.  The latter is retrieved dynamically by minimising the
spectral action.

Therefore, twisting the grand algebra down to the Standard
Model produces results similar to the ones of
\cite{Chamseddine:2013fk}. This raises questions on the interest of
the twist. As explained  in section
\ref{sec:twist-loretnz}, there is an
added value in twists, even if  not the one expected!
But before coming to that, let us try to generalize the twisting of the
grand algebra to arbitrary spectral triples.

\section{Minimal twist}
\label{sec:minimal-twist}

\subsection{Twisting up}
\label{sec:twisting-up}

The algebra
 ${\cal B}$ is not invariant under the twisting automorphism $\rho$
 because the grand algebra has been only partially twisted: only the quaternionic sector acts non-trivially on the
chiral index $s$. If one also makes the complex sector non trivial
on the chiral index, then the grading condition breaks the grand
algebra to
\begin{equation}
\left(\HHH_L^l \oplus\HHH_L^r \oplus \HHH_R^l\oplus \HHH_R^r\right) \oplus
 \left(M_4^l(\C)\oplus M_4^r(\C)\right),
\label{eq:53}
\end{equation}
which is invariant under $\rho$. This is twice the left-right algebra $\A_{LR}$ of \S
\ref{sec:grand-algebra}, which is broken to the algebra $\A_{SM}$ of the
Standard Model by the first-order condition of $\gamma^5\otimes D_F$.

 This suggests another approach to  twisting the
 Standard Model while preserving the first-order condition. Rather than twisting down a bigger algebra to 
 $\A_{SM}$, one may double $\A_{SM}$ to
 \begin{equation}
 \A_{SM}\otimes\C^2\simeq \A_{SM}\oplus\A_{SM},\label{eq:30}
 \end{equation}
then make each
 copy of $\A_{SM}$ act independently on the left/right components of
 spinors, and finally twist the commutator to avoid
 unboundedness problems.

This is a ``twisting up'' procedure, in which the idea is to use the
flexibility introduced by twisted spectral triples to enlarge the algebra -- hopefully
preserving the grading and the first-order conditions --  rather than 
using these conditions to constrain a bigger algebra.
The rule of the game is to leave the Hilbert space and the Dirac operator
untouched, in order not to  alter the fermionic content of the
model. As a side remark,  there exist
some models in noncommutative geometry that
introduce new fermions, as mentioned in the introduction, but since there
is  no phenomenological
indications of new fermions so far,  we limit ourselves to models that
preserve the fermionic sector. 

Given  a spectral triple
$(\A, \HH, D)$, the idea is thus to build a twisted spectral triple $(\A',
\HH, D), \rho$ with the same Hilbert space and Dirac operator, 
in such a way that the initial triple is retrieved as a
``non-twisted'' limit of the twisted one.
This led in \cite{Lett.} to define  the \emph{minimal twist}  of a spectral triple
$(\A, \HH, D)$ by a unital algebra $\cal B$ as a twisted spectral triple
$(\A\otimes {\cal B}, \HH, D), \rho$ such that the representation of $\A\otimes\I_{\cal B}$
coincides with the initial representation of $\A$. 

One may think of
other ways to implement the idea of ``non-twisted limit'', for
instance by simply asking that $\A'$ contains $\A$  as a subalgebra invariant under the
twist. More elaborate procedure  for untwisting a twisted spectral triple have been
proposed, for instance  in \cite{Goffeng:2019aa, Brzezinski:2018aa}.

An advantage of minimal twists is to allow to play with the
Standard Model, remaining close to it. For
almost commutative geometries -- i.e. the product of a manifold by a finite
dimensional spectral triple as in (\ref{eq:2}) -- then the only possible minimal
twist by a finite dimensional algebra is with ${\cal B}= \C^l\otimes
\C^2$, with $\rho$ the flip automorphism of $\C^2$ and $l\in\mathbb N$ a measure
of the non irreducibility of the representation of $\A_F$ on $\HH_F$
\cite[Prop. 4.4]{Lett.}.

 \subsection{Twist by grading}
\label{sec:twist-grading}

The twisting up procedure is easily  applicable to any
graded spectral triple $(\A, \HH, D)$. Indeed,  by definition, the grading $\Gamma$ commutes
with the representation of $\A$, so the latter actually is the direct sum of
two independent -- commuting -- representations of $\A$ on the eigenspaces $\HH_+$, $\HH_-$ of
$\Gamma$,
\begin{equation}
  \label{eq:38}
  \pi_+(a) = \frac 12\left(\I + \Gamma \right)a,\quad  \pi_-(a) = \frac 12\left(\I - \Gamma \right)a.
\end{equation}
In other words, decomposing $\HH$  as the
sum of the two  eigenspaces of $\Gamma$, the representation of
$\A$ is block diagonal. Thus there is enough space on $\HH$ to represent
$\A\otimes \C^2$ as
\begin{equation}
\pi((a, a'))=  \pi_+(a) +  \pi_-(a') \quad \forall (a, a')\in \A\otimes\C^2.
\label{eq:54}
\end{equation}
Let 
\begin{equation}
  \label{eq:48}
  \rho((a, a'))=(a', a) \quad \forall (a, a')\in \A\otimes\C^2
\end{equation}
denote the flip automorphism. Then the triple
\begin{equation}
(\A\otimes \C^2, \HH, D), \rho
\label{eq:55}
\end{equation}
with representation (\ref{eq:54}) is a graded twisted spectral triple \cite[Prop. 3.8]{Lett.}. In
addition, if the initial triple is real with real structure $J$, then
the latter  is also a real structure for the twisted spectral triple (\ref{eq:54}). In
particular the twisted first-order condition is automatically
satisfied.

This  \emph{twist by grading}  procedure  associates a twisted
 partner to any graded spectral triple, preserving a first-order
 condition. This seems the ideal way to twist the Standard Model. 
Unfortunately, this does not generate the extra scalar field. 
Indeed, one has that $\Gamma_F$ anticommutes independently with  $D_Y$
and $D_M$ (see e.g. \cite[\S 4.1]{Devastato:2013fk} for the computation in tensorial notations) so
in particular $\gamma^5\otimes D_M$ anticommutes with
$\Gamma=\gamma^5\otimes\Gamma_F$. This means that
\begin{eqnarray}
  \label{eq:56}
(\gamma^5\otimes D_M) \pi_+(a) = \pi_-(a)(\gamma^5\otimes D_M) +\frac
12(\I - \Gamma)[\gamma^5\otimes D_M,a ],\\
(\gamma^5\otimes D_M) \pi_-(a) = \pi_+(a)(\gamma^5\otimes D_M) +\frac
12(\I + \Gamma)[\gamma^5\otimes D_M,a ].
\end{eqnarray}
So
\begin{eqnarray}
\nonumber
  [\gamma^5\otimes D_M, \pi((a, a'))]_\rho&=  (\gamma^5\otimes D_M) (\pi_+(a)
  + \pi_-(a')) -  (\pi_+(a')
  + \pi_-(a)) (\gamma^5\otimes D_M),\\
  \label{eq:57}
&=[\gamma^5 \otimes D_M, a] + [\gamma^5 \otimes D_M, a'].
\end{eqnarray}
The right hand side is zero since 
$\gamma^5\otimes D_M$ commutes with the representation of
$\A$. Therefore
$\gamma^5\otimes D_M$ twist-commutes with the representation of $\A\otimes\C^2$. Hence the twist by grading  does not modify the situation: 
$\gamma^5\otimes D_M$ is transparent under under twisted fluctuations,
just like it was under usual fluctuations.

\subsection{Twisted fluctuation without the first-order condition}
\label{sec:twist-fluct-with}
 
The twist by grading is not the only possibility for twisting up the
Standard Model. As explained in \cite[below Prop.3.8]{Lett.}, in order
to minimally twist a spectral triple $(\A, \HH, D)$ by $\C^2$, one may
repeat the construction of the precedent section using, instead of
the grading~$\Gamma$,  any operator $\tilde\Gamma$ that
\begin{itemize}
\item squares to $\I$ and commutes with $\A$: this condition is 
  sufficient to guarantee that $\pi_+, \pi_-$ in
  (\ref{eq:38}) are two representations
  of $\A$ commuting with each other, and it becomes necessary as soon
  as $\A$
  is unital;

\item  is selfadjoint: this is to  guarantee that $\pi_+$ and $\pi_-$ are involutive representations;

\item  has both eigenvalues
  $+1, -1$ of non-zero multiplicity, so that neither $\pi_+$ nor
  $\pi_-$ is zero.
\end{itemize}
But there is no need for $\tilde\Gamma$ to anticommute with the
Dirac operator. This means that $\tilde\Gamma$ is not necessarily a
grading for the spectral triple.

A classification of all such twisting operators $\tilde\Gamma$ for almost
commutative geometries is on its way \cite{Filaci:2022vc}: the conditions necessary to make
the construction work actually reduce to a couple of relations on
$D_F$ only. The 
anticommutation with the Dirac operator seems to be required to
have the twisted first-order condition (but this has yet to be proved
in full generality). This would imply that the extra scalar field
and the twisted first-order condition be mutually exclusive.

Therefore it becomes relevant to extend to the twisted case the results of
\cite{Chamseddine:2013fk} regarding inner fluctuations without the
first-order condition. This has been done in
\cite{Martinetti:2021aa}, where it was shown that the removal of the
twisted first-order condition yields a second order term in the
twisted fluctuation (\ref{eq:24}), which is a straightforward adaptation of the term
worked out in the non-twisted case.

Following this path, a minimal twist of the Standard Model has been
worked out in great details in \cite{M.-Filaci:2020aa}, that does not
preserve the twisted first-order condition and generates the extra
scalar field. The gauge part of this model is similar to the Standard
Model's, and the Higgs sector is made of two Higgs doublets which are
expected to combine in a single doublet in the action. There is the extra scalar
field with two components $\sigma_l$, $\sigma_r$ acting independently
on the chiral components of spinors, and finally, there is also an unexpected new 
field of $1$-forms $X_\mu$, whose interpretation is discussed in the next section.

 \section{Twist and change of signature} 
 \label{sec:twist-loretnz}

At this point of our journey through twisted spectral triples, one seems
to be back
to the starting point: twisted spectral triples solve the unboundeness of the commutator of the grand
algebra with
$\ds\otimes\I$, but they do not permit to generate the extra scalar
field, unless one violates the twisted first-order condition. What is
then their  added value?

The interest of the twist is not so much in the generation of
the extra scalar field than in the new field of $1$-form $X_\mu$ mentioned above. 
This field was already observed in
\cite{buckley}, and its appearance actually does not  depend on the details of
the model \cite{Martinetti:2014aa}: it is
intrinsic to  minimal twists of almost
commutative geometries. Even in the simplest case of a 
minimally twisted four dimensional manifold (without any product by a finite dimensional
structure), a twisted fluctuation of
the Dirac operator $\ds$ yields a field of $1$-forms, 
in contrast  with the
non twisted case where $\ds$ does not fluctuate.

The physical
sense of this fluctuation remained obscure, until
it was confronted with an observation made in \cite{Devastato:2018aa}:
a twist induces on the Hilbert space a new
 inner product with \emph{Lorentzian} \emph{signature}. Furthermore,
 this product permits to define a twisted version of the fermionic
 action. In some example detailed below, in this action formula the
 field $X_\mu$ identifies with the (dual of) the $4$-momentum
in Lorentzian signature \cite{Martinetti:2019aa}.

\subsection{Twisted inner product}
\label{sec:twist-inner-prod}

A gauge transformation  (\ref{eq:20}),
$D_A \to \textrm{Ad}(u) \,D_A\,
\textrm{Ad}(u)^{-1}$,
preserves the selfadjointness of
the covariant Dirac operator $D_A$, for $\textrm{Ad}(u)^{-1}=Ju^*J^{-1}u^*=\textrm{Ad}(u)^*$. A twisted gauge~transformation~(\ref{eq:58}) 
\begin{equation}
  \label{eq:61}
  D_{A_\rho} \to \rho(\textrm{Ad}(u)) \,D_{A_\rho}\,
\textrm{Ad}(u)^{-1}
\end{equation}
does not. Is there some selfadjointness which is preserved by
(\ref{eq:61})?

 There is a natural
inner product associated with a twisted spectral triple, as soon as
the twisting autormorphism $\rho$ extends to an inner automorphism of ${\cal B}(\HH)$:
\begin{equation}
\rho({\cal O}) = R{\cal O} R^\dag \qquad \forall {\cal O}\in{\cal
  B}(\HH)
\label{eq:63}
\end{equation}
for some unitary operator $R$ on $\HH$. Namely, the 
$\rho$-inner product \cite{Devastato:2018aa}
\begin{equation}
\langle\Psi,\Phi\rangle_\rho:=\langle\Psi,
  R\Phi\rangle.\label{eq:64}
\end{equation}
Since
 $ \langle\Psi,\mathcal{O}\Phi\rangle_{\rho}=\langle\rho(\mathcal{O})^\dag\Psi,\Phi\rangle_{\rho}$,
the adjoint of $\cal O$ with respect to this new product is 
\begin{equation}
{\cal O}^+:=\rho({\cal O})^\dag.
\label{eq:62}
\end{equation}
 
If the unitary $R$ commutes or anticommutes with the real structure,
then $\rho(\textrm{Ad}(u))$ as defined before (\ref{eq:59}) coincides with
$R\textrm{Ad}(u)R^*$ (making the notation $\rho(\textrm{Ad}(u))$ 
unambiguous). In addition, 
\begin{eqnarray}
 \left( \textrm{Ad}(u)^{-1}\right)^+= \left(R  J u^* J^{-1}u^*R^*\right)^\dagger =RuJuJ^{-1}R^*=\rho(\textrm{Ad}(u)).
\end{eqnarray}
Therefore  a twisted gauge transformation (\ref{eq:61}) preserves the selfadjointness
with respect to the $\rho$-inner product.

\noindent{\bf Example}: The minimal twist of a 
Riemannian spin manifold $\M$ of even dimension $2m$~is
\begin{equation}
   \label{eq:18400}
\A= \cinf\otimes\,\C^2, \quad \HH= L^2(\M,S), \quad  D=\ds;  \quad \rho
 \end{equation}
with twisting automorphism the flip  $\rho(f,g) = (g,f)$ for $f, g$ in
$\cinf$. The representation~is 
 \begin{equation}
   \pi(f,g)=\left(\begin{array}{cc} f\,\I_{2^{m-1}}& 0 \\  0&
                                                              g\I_{2^{m-1}}\end{array}\right)\qquad
                                                          \forall (f,g)\in
\A.
\label{eq:18700}
\end{equation}
The flip  $\rho$ extends to
the inner automorphism of ${\cal B}(\HH)$ that exchanges the element
on the diagonal and on the off-diagonal, implemented for instance by
$R =\gamma^0$ the first Dirac matrix.
Then the $\rho$-product (\ref{eq:64})  
\begin{equation}
  \label{eq:65}
 \langle\Psi,\Phi\rangle_\rho=\int_\M \Psi^\dag \gamma^0 \Phi\, d^4x 
\end{equation}
coincides pointwise with the Krein product for the
  space of spinors on a \emph{Lorentzian manifold} (only pointwise,
  for the manifold on which one integrates is still Riemannian).

This example points towards a link between twists and a kind of
transition from Euclidean to Lorentzian signatures: by fluctuating a
twisted Riemannian manifold, one ends up preserving a
Lorentzian product!  However, the twist is not an 
implementation of Wick rotation in noncommutative geometry (for this,
see \cite{DAndrea:2016aa}): a twisted fluctuation (\ref{eq:61}) does not turn the 
operator $D_{A_\rho}$,  selfadjoint for the initial
(Euclidean) inner
product, into an operator $D_{A_\rho^u}$ selfadjoint for the
Lorentzian product.{\footnote{If one were starting with an operator selfadjoint for the
twisted product, much in the vein of \cite{Dungen:2015aa}, then  this
selfadjointness would be preserved by
twisted fluctuation.}}  A better understanding of the link between
twist and Lorentzian signature follows from the  study of the
fermionic action.

\subsection{Fermionic action}
\label{sec:fermionic-action}

Given a real spectral triple $(\A, \HH, D)$, the fermionic action for
the covariant operator $D_A$ is \cite{Chamseddine:2007oz}
\begin{equation}
S^f(D_A)= \frak A_{D_A}(\tilde \xi, \tilde \xi)\label{eq:66}
\end{equation}
with $\tilde \xi$ the Grassman variables associated to $\xi\in\HH^+=\left\{\xi\in\HH, \Gamma\xi =\xi\right\}$
and
\begin{equation}
\frak A_{D_A}(\xi, \xi') =\langle J\xi, D_A\xi'\rangle\label{eq:67}
\end{equation}
the antisymmetric bilinear form defined by $D_A$ and the
real structure $J$. The latter is needed to make the form
antisymmetric (hence applicable on Grassman variables). 
One restricts to the eigenspace $\HH^+$ of the grading because of the fermion doubling~\cite{fermiondoublingNaples1}. This also  makes sense physically,
  for  $\HH^+$ is the subspace of 
  $\HH$ generated by the elements
  $\psi\otimes\Psi$ with a well defined chirality (that
  is $\psi\in L^2(\M, S)$ and  $\Psi\in \HH_F$ are eigenvectors of $\gamma^5$, $\Gamma_F$ with the
  same eigenvalue).

For a twisted spectral triple $(\A, \HH,D), \rho$ as in \S 
\ref{sec:twist-inner-prod}, the fermionic action is \cite{Devastato:2018aa}
\begin{equation}
S^f(D_{A_\rho})= \frak T_{D_{A_\rho}}(\tilde \xi,
\tilde\xi) \label{eq:68}
\end{equation}
for $\xi\in\HH_r:=\left\{\xi\in\HH, R\xi=\xi\right\}$, $\tilde{.}$
the Grassmann variables and
$$
\frak T_{D_{A_\rho}}(\xi, \xi') := \langle J\xi, RD_{A_\rho}\xi'\rangle.
$$
One inserts the matrix $R$ in the bilinear form in order to make the
action (\ref{eq:68}) invariant under
a twisted gauge transformation (\ref{eq:58}) (the same is true in case
there is no first-order condition \cite{Martinetti:2021aa}). 
The restriction to $\HH_r$ guarantees that the bilinear form
be antisymmetric.

\subsection{Twisted fluctuation as Lorentzian $4$-momentum}
\label{sec:twist-fluct-as}

We begin with the minimal twist
(\ref{eq:18400}) of a $4$-dimensional manifold. The $+1$ eigen\-space
of $R=\gamma^0$ is spanned by Dirac spinors of the form
$\xi=\left(\begin{array}{c}\zeta\\ \zeta\end{array}\right)$ with
$\zeta$ a Weyl spinor.
A selfadjoint twisted fluctuation~(\ref{eq:24}) sends $\ds$ to the covariant
operator
\begin{equation}
\ds_{A_\rho}= \ds -i\, X_\mu\gamma^\mu,
\label{eq:71}
\end{equation}
 parametrised by the $1$-form field
 \begin{equation}
   X_\mu =f_\mu\gamma^5\quad \textrm{ with } \quad f_\mu\in C^{\infty}(\M, \mathbb R).
 \end{equation}
The twisted  fermionic action  is \cite[Prop. 3.5]{Martinetti:2019aa}
\begin{equation}
S^f(\ds_{A_\rho}) = 2\int_\M d\mu\;  \bar{
\tilde \zeta}^\dagger\sigma_2\,(if_0-\sum_{j=1}^3\sigma_j\partial_j)\, \tilde
\zeta.
\end{equation}

The integrand reminds of the Weyl Lagrangian -- which
lives in Lorentzian signature
\begin{equation}
i\psi_l^\dag\,
  \tilde\sigma_M^\mu\, \partial_\mu \psi_l\qquad\textrm{ where }\quad
\tilde\sigma_M^\mu :=\left\{\mathbb I_2,
  -\sigma_j\right\},\label{eq:70}
\end{equation}
except that  the $\partial_0$ derivative is missing. It can be
restored 
assuming that
$\zeta$ is a plane wave function of energy $f_0$ (in unit $\hbar
=1$) with spatial part $\zeta(\bf x)$, that is
\begin{equation}
\zeta (x_0, {\bf x})=
e^{if_0x_0}\zeta(\bf x).
\label{eq:75}
\end{equation}
Then the integrand reads (modulo an irrelevant factor $2$) as $\bar{\tilde \zeta}^\dagger\sigma_2\,\tilde\sigma^\mu_M \partial_\mu\, \tilde\zeta$. 
However, this cannot be identified with the Weyl Lagrangian
(\ref{eq:70}) because of the
extra $\sigma_2$ matrix which forbids the simultaneous identification
of $\tilde\zeta$ with $\psi_l$ and $\bar{
\tilde \zeta}^\dagger\sigma_2$ with $i\psi_l^\dagger$. 
 In other terms, there are not enough degrees of freedom to identify the
fermionic action of a twisted manifold with the Weyl Lagrangian. 
\medskip

This can be cured by doubling the manifold. Namely one considers the product 
 \begin{equation}
(\cinf\otimes
 \C^2, L^2({\cal M,S}) \otimes \mathbb{C}^2, \ds \otimes
 \mathbb{I}_2).
\label{eq:79}
 \end{equation}
of $\M$ by a finite dimensional spectral
 triple $(\C^2, \C^2, 0)$.
Its minimal twist is
 \begin{equation}
{\cal A} = \left(C^\infty({\cal M}) \otimes \mathbb{C}^2\right) \,
\otimes \,\C^2, \quad
	{\cal H} = L^2({\cal M,S}) \otimes \mathbb{C}^2, \quad
	D = \ds \otimes \mathbb{I}_2
\label{eq:73}
              \end{equation}
with representation 
\begin{equation}
	\pi(a, a') =
	\left(\begin{array}{cccc}
		f\mathbb{I}_2 & 0 & 0 & 0 \\ 0 & f'\mathbb{I}_2 & 0 & 0 \\ 
		0 & 0 & g'\mathbb{I}_2 & 0 \\ 0 & 0 & 0 & g\mathbb{I}_2
	\end{array}\right)\quad  a=(f,g), \, a'=(f', g') \in \A
\label{eq:74}  
    \end{equation}
and twist $\rho(a, a')=(a', a)$. The latter is implemented by the unitary $R=\gamma^0\otimes \mathbb I_2$,
 whose $+1$ eigenspace ${\HH_r}$ is now spanned by $\left\{\xi\otimes e, \phi\otimes \bar e\right\}$ where $\left\{e, \bar e\right\}$
is a basis of  $\C^2$ and
  \begin{equation}
\xi =\left( \begin{array}{c} \zeta\\ \zeta \end{array}\right), \quad\phi
  =\left(\begin{array}{c} \varphi\\ \varphi \end{array}\right)\label{eq:77}
  \end{equation}
are Dirac spinors with $\zeta$ and $\varphi$ Weyl spinors.
A selfadjoint twisted fluctuation  of $D$,
\begin{equation}
  \label{eq:76}
D_{A_\rho}=  D -iX_\mu\gamma^\mu\otimes\I_2 + g_\mu\gamma^\mu\otimes \Gamma_F
\end{equation}
with $\Gamma_F$  the grading of the finite dimensional spectral triple
\cite[Prop. 4.3]{Martinetti:2019aa}, is parametrised
by the same field $X_\mu$ as before and a second $1$-form field
\begin{equation}
  g_\mu\I_4 \quad\textrm{with} \quad g_\mu\in\cinf.
\end{equation}

For a vanishing $g_\mu$,
the fermionic action is the integral of \cite[Prop. 4.4]{Martinetti:2019aa}
\begin{equation}
{\cal L} := 4\bar{\tilde\varphi}^\dag\sigma_2 
	\left(if_0 - \textstyle\sum_{j=1}^3\sigma_j\partial_j \right)
        \tilde\zeta.\label{eq:78}
              \end{equation}
One retrieves the Weyl Lagrangian (\ref{eq:70}) by identifying the physical Weyl spinors as  
$\psi_l := \tilde\zeta$ and  $\psi_l^\dag := 
	-i\bar{\tilde\varphi}^\dag\sigma_2$, then assuming $\psi_l$ be of
the form (\ref{eq:75}), that is $\partial_0\psi_l =if_0\psi_l$. 
   Thus the fermionic action for a twisted doubled
   Riemannian manifold describes a plane wave solution of Weyl
   equation, in  Lorentzian signature, whose $0$\textsuperscript{th} component
   of the 4-momentum is $p_0=-f_0$.
The  result also holds for the right-handed Weyl equation (see
\cite[Prop. 4.5]{Martinetti:2019aa}).
   \medskip

A similar analysis holds for the spectral triple of electrodynamics
proposed in \cite{Dungen:2011fk}. Its minimal twist is
\begin{equation*}
{\cal A}_{ED}= \left(C^{\infty}(\mathcal{M}) \otimes
\,\,\mathbb{C}^2\right)\, \otimes \C^2,\;  {\cal H} = L^2(\mathcal{M,S}) \otimes \mathbb{C}^4,
	\quad D = \ds\otimes \mathbb{I}_4 + \gamma^5 \otimes
        D_{\cal F}\label{eq:69}
  \end{equation*}
where the internal Dirac operator and the representation are 
$$D_{\cal F} =
\left(\begin{array}{cccc}
	0 & d & 0 & 0 \\ \bar{d} & 0 & 0 & 0 \\ 0 & 0 & 0 & \bar{d} \\ 0 & 0 & d & 0
\end{array}\right),\; 	\pi(a,a') =
	\left(\begin{array}{ccccccccc}
		f\mathbb{I}_2 & 0 & 0 & 0 & 0 & 0 & 0 & 0 \\
		0 & f'\mathbb{I}_2 & 0 & 0 & 0 & 0 & 0 & 0 \\
		0 & 0 & f'\mathbb{I}_2 & 0 & 0 & 0 & 0 & 0 \\
		0 & 0 & 0 & f\mathbb{I}_2 & 0 & 0 & 0 & 0 \\
		0 & 0 & 0 & 0 & g'\mathbb{I}_2 & 0 & 0 & 0 \\
		0 & 0 & 0 & 0 & 0 & g\mathbb{I}_2 & 0 & 0 \\
		0 & 0 & 0 & 0 & 0 & 0 & g\mathbb{I}_2 & 0 \\
		0 & 0 & 0 & 0 & 0 & 0 & 0 & g'\mathbb{I}_2
	\end{array} \right) 
$$
with $d\in \C$, $a=(f,g)$, $a'=(f', g')$ in $\cinf\otimes \C^2$. The
twist $\rho(a, a')= (a', a)$ extends to an inner automorphism of
${\cal B}(\HH)$ generated by the unitary $\gamma^0\otimes\I_4$. Its
$+1$-eigenspace is generated by
\begin{equation}
  \label{eq:80}
  \xi_1\otimes e_l, \quad\xi_2\otimes e_r, \quad \phi_1\otimes \overline{e_l},\quad \phi_2\otimes \overline{e_r}, 
\end{equation}
where $\xi_k$, $\phi_k$ ($k=1,2$) are Dirac spinors of the form (\ref{eq:77}) 
while $\left\{ e_l, e_r, \overline{e_l}, \overline{e_r}\right\}$ is an
orthonormal basis of $\C^4$. 

A selfadjoint twisted fluctuation of $D$ is parametrized by the same two
$1$-form fields as before \cite[Prop. 5.3]{Martinetti:2019aa}
\begin{equation}
  \label{eq:81}
  D_{A_\rho}= D -i X_\mu\gamma^\mu\otimes \I' +   g_\mu\gamma^\mu\otimes \I''
\end{equation}
where $\I'=\textrm{diag}(1,-1,1,-1)$, $\I''=\textrm{diag}(1,1,-1,-1)$
(the part $\gamma^5\otimes D_F$ is
transparent under twisted fluctuation: there is no Higgs field in
classical electrodynamics!).
Under a  gauge transformation (\ref{eq:58}), one has that $f_\mu$ is invariant
while $g_\mu$ trasforms as the $U(1)$ gauge potential of
electrodynamics.

The spectral action is the integral of \cite[Prop. 5.12]{Martinetti:2019aa}
\begin{equation}
  \label{eq:83}
\vspace{-.5truecm}  {\cal L}_\rho^f=\bar{\tilde\varphi}_1^\dagger\sigma_2\left(if_0-\sum_j\sigma_j{\cal
      D}_j\right)\tilde\zeta_1-
\bar{\tilde\varphi}_2^\dagger\sigma_2\left(if_0+\sum_j\sigma_j{\cal D}_j\right)\tilde\zeta_2
+ \left(\bar d\bar{\tilde \varphi}^\dagger_1\sigma_2\bar{\zeta}_2 +  d\bar{\tilde \varphi}^\dagger_2\sigma_2\bar{\zeta}_1\right)\end{equation}
where ${\cal D}_\mu=\partial_\mu - ig_\mu$ is the covariant derivative
associated to the electromagnetic $4$-potential.
Defining the physical spinors as
\begin{equation}
  \label{eq:84}
  \psi=\left(
    \begin{array}{c}
      \psi_l\\ \psi_r
    \end{array}\right):=\left(
    \begin{array}{c}
      \tilde\zeta_1\\ \tilde\zeta_2
    \end{array}\right),\quad
\psi^\dagger=\left( \psi_l^\dagger,
  \psi_r^\dagger\right):=\left(-i\bar{\tilde\varphi}_1^\dagger\sigma_2, i\bar{\tilde\varphi}_2^\dagger\sigma_2\right)
\end{equation}
then assuming that $\partial_0\psi= if_0\psi$ and imposing $d=-im$ with
$m>0$ to be purely imaginary, the Lagrangian
(\ref{eq:83}) reads
\begin{equation}
  \label{eq:85}
  {\cal L}_M=i\psi_l^\dagger\left({\cal
      D}_0-\sum_j\sigma_j{\cal
      D}_j\right)\psi_l+i\psi_r^\dagger \left({\cal
      D}_0+\sum_j\sigma_j{\cal
      D}_j\right)\psi_r - m \left(\psi_l^\dagger\psi_r + \psi_r^\dagger\psi_l\right).
\end{equation}
 This is the Dirac Lagrangian in Minkowski spacetime, for a mass $m$,
 in the temporal gauge (that is ${\cal D}_0 =\partial_0$). Hence the
 fermionic action for the minimal twist of the spectral triple of
 electrodynamics describes a plane wave solution of the Dirac equation
 in Lorentz signature, with $0$\textsuperscript{th} component of the
 4-momentum $p_0=-f_0$. 

By implementing the action of boosts on $L^2(\M,S)\otimes \C^2$, one
is able to identify the other components of the fluctuation $f_\mu$
with the other components of the $4$-momentum. However this is
obtained at the level of the equation of motion, not for the
Lagrangian density (see \cite[\S 6.1]{Martinetti:2019aa}).

\section{Conclusion and outlook}
\label{sec:conclusion-outlook}

The idea of using twisted spectral triples in high-energy physics was
born with the hope of generating the extra scalar field needed to stabilise
the electroweak vacuum (and to fit the Higgs mass), respecting the axioms of noncommutative
geometry. More specifically it was thought that the first-order
condition could be twisted, rather than abandoned.  We have shown in
this note that this is not possible. This moves the interest of the
twist towards what seemed at first sight a side effect, namely the
non-zero twisted fluctuation of the free Dirac
operator $\ds$. It yields a new field of $1$-forms, whose physical
meaning becomes clear by computing the fermionic action. For the minimal twist of a doubled
manifold, and the minimal twist of the spectral triple of
electrodynamics, this fields identifies with (the dual of) the $4$-momentum in Lorentzian signature. Preliminary computations indicate that a similar result also holds for the
 twist of the Standard Model presented in \cite{M.-Filaci:2020aa}.

It remains to understand why one ends up in the temporal gauge and,
more importantly, if the identification between twisted fluctuation
of $\ds$ and the $4$-momentum still
makes sense  for the bosonic part of the action, given by the spectral
action.  Not to mention that the definition of the latter in a twisted
context has not been stabilised yet \cite{Devastato:2020aa}.

\section*{Acknowledgments}
This research is part of the project No. 2021/43/P/ST1/02449 co-funded by the National Science Centre
and the European Union's Horizon 2020 research and innovation
programme under the Marie Sk\l odowska-Curie
grant agreement no. 945339. For the purpose of Open Access, the author has applied a CC-BY public copyright
licence to any Author Accepted Manuscript (AAM) version arising from this submission.

\noindent It is part of the second author's activity in the
mathematical physics group of INDAM.

\end{document}